\documentclass[12pt]{article}
\usepackage{latexsym}
\usepackage{amsmath}
\def\d{\delta}
\def\e{\epsilon}

\def\g{\gamma}

\def\l{\lambda}
\def\m{\mu}
\def\n{\nu}

\def\s{\sigma}

\def\S{\Sigma}

\def\be{\begin{equation}}
\def\ee{\end{equation}}
\def\arr{\begin{array}{rll}}
\def\ea{\end{array}}
\def\bea{\begin{eqnarray}}
\def\eea{\end{eqnarray}}

\begin{document}
\renewcommand{\thefootnote}{\fnsymbol{footnote}}
\begin{titlepage}
\setcounter{page}{0}
\begin{flushright}
LMP-TPU--11/08  \\
\end{flushright}
\vskip 1cm
\begin{center}
{\LARGE\bf
Quartet unconstrained formulation}\\
\vskip 0.5cm
{\LARGE\bf for massive higher spin fields}\\
\vskip 2cm
$
\textrm{\Large I.L. Buchbinder~${}^{a}$, A.V. Galajinsky~${}^{b}$\ }
$
\vskip 0.7cm
${}^{a}${\it Department of Theoretical Physics, Tomsk State Pedagogical University, \\
634041 Tomsk, Russian Federation} \\
{Email: joseph@tspu.edu.ru}
\vskip 0.7cm

${}^{b}${\it
Laboratory of Mathematical Physics, Tomsk Polytechnic University, \\
634050 Tomsk, Russian Federation} \\
{Email: galajin@mph.phtd.tpu.edu.ru}

\end{center}
\vskip 1cm
\begin{abstract} \noindent
We generalize the unconstrained description of free
massless higher spin fields pre\-viously developed in [Nucl.Phys. B 779 (2007) 155] to the case
of free massive higher spin fields in
a flat space of arbitrary dimension. The Lagrangian
is given in an easy-to-handle form for an arbitrary value of spin. It is
local, free from higher derivative terms, and involves a minimal
number of auxiliary fields needed for
an unconstrained gauge invariant description of a free massive higher spin
field in arbitrary dimension.
\end{abstract}

\end{titlepage}

\renewcommand{\thefootnote}{\arabic{footnote}}
\setcounter{footnote}0

\vspace{0.5cm}

\noindent
{\bf 1. Introduction}\\

Last three decades have seen interesting evolution of the Lagrangian description of free massive higher spin fields in flat space and on
anti de Sitter background. In the original works of Singh and Hagen \cite{sh,sh1} a massive spin--$s$ boson was described in terms of
a totally symmetric traceless tensor field of rank--$s$, while a massive spin--$(n+\frac 12)$ fermion was represented by
a totally symmetric $\g$--traceless spin--tensor of rank--$n$. A peculiar feature of the Singh--Hagen formalism is that, in order to
derive correct equations of motion from a Lagrangian, one needs to introduce auxiliary fields, the fact anticipated by Fierz and Pauli
long ago in \cite{fp}. The auxiliary fields are traceless and enter the Lagrangian with specific number coefficients. The procedure of
fixing the number coefficients is tedious and their final form is very complicated. The aforementioned trace conditions imply
that the theory is given in terms of off-shell constrained fields.

The formulations proposed in \cite{zin,rus1} (for a related work see \cite{rus,zin1})
provide an interesting alternative to the Singh--Hagen formalism and generalize the latter to the case of anti de Sitter background.
The principal new ingredient is
the gauge symmetry of the massive higher spin field Lagrangian which facilitates computation of the Singh--Hagen
coefficients.
Auxiliary fields play the role of
Stueckelberg fields which can be gauged away leaving one with a single massive spin--$s$ mode.
As the gauge invariant formulations are constructed starting from the massless higher spin theories \cite{FR,FF},
they share with the latter the trace constraints on physical and
auxiliary fields as well as those on
gauge parameters \cite{zin,rus1}. Thus, these formulations are also given in terms of off-shell constrained fields.

A completely unconstrained description for massive higher spin
fields in flat space and on anti de Sitter background was achieved
within the universal BRST approach \cite{bk1}--\cite{mr} (see also the review \cite{tf}).
Here a spin--$s$ field is represented by a vector in an
appropriate Fock space and the conditions which determine an
irreducible massive representation of the Poincar\'e group or
the anti de Sitter group
come about as operators annihilating the state. Treating these
operators as constrains one can derive the canonical BRST charge.
The action functional is then constructed in terms of the BRST
charge with the use of a special technique
(see \cite{bk1}--\cite{mr} for more details).

Although the BRST approach produces Lagrangian formulations in terms
of unconstrained fields and gauge parameters, it is very
general. A lot of auxiliary fields enter the formulation. The resulting gauge theory is reducible
and in the case of massive higher spin fermionic fields the order of
reducibility grows with the value of spin. Besides, for an arbitrary
value of spin an explicit form of the action functional in terms of
space-time fields (not the Fock space vectors) has not yet been derived.

Quite recently, the geometric approach to unconstrained description of massless higher spin fields
developed in \cite{fs1}--\cite{fms} was generalized to the case of massive higher spin bosons in flat space and on
anti de Sitter background \cite{fr,fms1}. The
resulting formulations are either nonlocal or involve higher derivatives acting on auxiliary fields.
In principle, the higher derivative terms can be eliminated by introducing extra auxiliary fields. For massless
higher spin bosonic and fermionic fields in flat space and on anti de Sitter background this was demonstrated in
\cite{bgk} (see also \cite{fr} for the case of massless bosons in flat space).

The purpose of this work is to construct an easy-to-handle unconstrained gauge invariant Lagrangian
formulation for free massive higher
spin fields in flat space of arbitrary dimension which unifies in a nice way the advantages of the BRST formulation
and the geometric approach.
It can be viewed as a consistent truncation of the Lagrangians obtained within the
BRST method which aims to keep a number of auxiliary fields at a reasonable minimum. At the same time, from the very beginning,
it possesses all the standard attributes of a classical field theory like locality, the absence of higher derivative terms etc.

For massless higher spin bosonic and fermionic fields in flat space and on anti de Sitter background such a formulation has
been constructed recently in \cite{bgk} (see also \cite{fr} for massless bosons in flat space).
It relies upon the so called triplet of fields \cite{fs,fs2,st} (for a frame-like description of the triplets in flat
space and on anti de Sitter background see \cite{sv}).
The triplet naturally accommodates
higher spin gauge symmetry with an unconstrained gauge
parameter. It describes a chain of irreducible spin modes and
admits a simple Lagrangian description \cite{fs,st}. In \cite{bgk}
unconstrained Lagrangian formulations for massless higher spin fields
in flat space and on anti de Sitter background
were systematically derived
from the triplets by finding an appropriate
set of gauge invariant constraints which extract a single
spin--s mode from the chain of irreducible representations contained in the triplet.
In order to write the
constraints without spoiling the unconstrained gauge symmetry, one
has to introduce an additional compensator. Ultimately, one arrives at a simple Lagrangian
formulation, which is local, free from higher
derivative terms and uses a quartet of fields for
an unconstrained description of any value of spin.

In this paper we generalize the quartet unconstrained formulation of \cite{bgk} to the massive case.
When describing massive higher spin bosons, it proves convenient to use dimensional reduction. So,
in the next section we briefly discuss the reduction mechanism we adhere in this work.
Sect. 3 is devoted to an unconstrained Lagrangian description of a massive spin--$s$ boson in a flat space.
After the dimensional reduction,
each member of the quartet gives rise to a collection of fields, including Stueckelberg fields.
The resulting formulation is given in an easy-to-handle form and enjoys irreducible
gauge invariance. In Sect. 4 we generalize the consideration to the fermionic case, this time without
making use of the dimensional reduction.  We summarize our results and discuss possible further
developments in the concluding Sect. 5.

\vspace{0.5cm}

\noindent
{\bf 2. Dimensional reduction}\\

In this section we fix the notation and discuss a dimensional reduction mechanism which will be used below.

Within the metric--like approach a spin--$s$ field is described by a totally symmetric
tensor of rank--$s$. Throughout the work the vector indices in $\mathcal{D}+1$ dimensions will be denoted by capital Latin letters, while those in
$\mathcal{D}$ dimensions by small Greek letters.

When analyzing equations of motion and gauge transformations for higher spin fields,
it proves convenient to switch to the notation which suppresses
vector indices and automatically takes care of symmetrizations. This is done
by introducing an auxiliary vector variable, say $Y^A$, such that
\be
\Phi_{A_1 \dots A_s} (X) \quad \Leftrightarrow \quad \Phi^{(s)}(X,Y)=\Phi_{A_1 \dots A_s} (X)~ Y^{A_1} \dots Y^{A_s}.
\ee
Here $X^A$ are coordinates parameterizing a $(\mathcal{D}+1)$--dimensional pseudo--Riemannian space--time with the
metric $\eta_{AB}=\mbox{diag}(+,-,+,\dots,+)$.
Denoting $P_A=\partial_A=\frac{\partial}{\partial X^A}$ and $\Pi_A=\frac{\partial}{\partial Y^A}$, one
has $P^2$ for the d'Alembertian, $\Pi^2$ for the trace, $(P\Pi)$ for the divergence and $(YP)$ for the derivative of a
field followed by symmetrization of indices
\bea
&&
P^2 ~\Phi^{(s)} (X,Y) \quad \quad \Leftrightarrow \quad \Box \Phi_{A_1 \dots A_s} (X)\ ,
\nonumber\\[2pt]
&&
\Pi^2 ~\Phi^{(s)} (X,Y) \quad \quad \Leftrightarrow \quad s(s-1) {\Phi^B}_{B A_1 \dots A_{s-2}} (X)\ ,
\nonumber\\[2pt]
&&
(P\Pi) ~\Phi^{(s)} (X,Y) \quad \Leftrightarrow \quad s \partial^B \Phi_{B A_1 \dots A_{s-1}} (X)\ ,
\nonumber\\[2pt]
&&
(YP) ~\Phi^{(s)} (X,Y) \quad \Leftrightarrow \quad \frac{1}{(s+1)} (\partial_{A_1} \Phi_{A_2 \dots A_{s+1}} (X)+\dots+\partial_{A_{s+1}} \Phi_{A_1 \dots A_s} (X))\ .
\eea

In passing from $\mathcal{D}+1$ to $\mathcal{D}$ dimensions, we follow a conventional recipe (see e.g. \cite{fms1}).
Both the physical and auxiliary coordinates are split
\be
X^A \quad \rightarrow \quad (x_0,x^\m)\ , \qquad Y^A \quad \rightarrow \quad (y_0,y^\m)\ ,
\ee
such that a tensor field of rank $s$ in $\mathcal{D}+1$ dimensions turns into a collection of fields of ranks
$s,s-1,\dots,0$ in $\mathcal{D}$
dimensions.
The metric in $\mathcal{D}$ dimensions reads $\eta_{\m\n}=\mbox{diag}(-,+,\dots,+)$.
The dependence of a resulting composite field on the Kaluza--Klein coordinate $x_0$ is fixed by the factor $e^{imx_0}$
\bea\label{fiel}
&&
\Phi^{(s)}(X,Y)=(\phi_{\m_1 \dots \m_s} (x) \cdot y^{\m_1} \dots y^{\m_s}+i \phi_{\m_1 \dots \m_{s-1}} (x) \cdot y^{\m_1} \dots y^{\m_{s-1}} \cdot y_0
+
\nonumber\\[2pt]
&& \quad
+\phi_{\m_1 \dots \m_{s-2}} (x)\cdot y^{\m_1} \dots y^{\m_{s-2}} \cdot y_0^2+i\phi_{\m_1 \dots \m_{s-3}} (x)\cdot y^{\m_1} \dots y^{\m_{s-3}} \cdot y_0^3+\dots)
e^{imx_0}=
\nonumber\\[2pt]
&& \quad
=(\phi^{(s)}(x,y)+i \phi^{(s-1)} (x,y) \cdot y_0+\phi^{(s-2)} (x,y)\cdot  y_0^2+i \phi^{(s-3)} (x,y) \cdot y_0^3+\dots)e^{imx_0},
\eea
where the real parameter $m$ is interpreted as the mass of each single component in $\mathcal{D}$ dimensions.
Notice that in our notation odd powers of the auxiliary variable $y_0$ are accompanied by the extra factor $i$.
It turns out that such a choice leads to reasonable real equations of motion for the component fields and yields a real Lagrangian.
In what follows, we use capital letters in order to designate composite fields like in (\ref{fiel}). Small letter are reserved for the components.

Denoting $p_\mu=\partial_\m=\frac{\partial}{\partial x^\m}$, $\pi_\m=\frac{\partial}{\partial y^\m}$, $\pi_0=\frac{\partial}{\partial y_0}$, one
can easily transport various operators from $\mathcal{D}+1$ to $\mathcal{D}$ dimensions. For example,
\bea
&&
P^2 \quad \quad \rightarrow \quad p^2-m^2\ , \quad \qquad \quad \Pi^2 \quad \quad \rightarrow \quad  \pi^2+\pi_0^2\ ,
\nonumber\\[2pt]
&&
(P\Pi) \quad \rightarrow \quad (p\pi)+i m \pi_0\ , \qquad
(YP) \quad \rightarrow \quad (yp)+i m y_0\ .
\eea

When constructing Lagrangians, it proves convenient to deal with a conjugate field (operator). This is obtained from (\ref{fiel})
by changing $y^\m \rightarrow \pi_\m$, $y_0 \rightarrow \pi_0$, which is followed by complex conjugation of the components
\bea\label{field}
&&
\hat{\bar\Phi}^{(s)}(X,\Pi)= \frac{1}{s!} e^{-imx_0}(\bar\phi^{\m_1 \dots \m_s} (x) \cdot \pi_{\m_1} \dots \pi_{\m_s}-i\bar\phi^{\m_1 \dots \m_{s-1}} (x) \cdot \pi_{\m_1} \dots \pi_{\m_{s-1}} \cdot \pi_0
+
\nonumber\\[2pt]
&& \quad
+\bar\phi^{\m_1 \dots \m_{s-2}} (x)\cdot \pi_{\m_1} \dots \pi_{\m_{s-2}} \cdot \pi_0^2-i\bar\phi^{\m_1 \dots \m_{s-3}} (x)\cdot \pi_{\m_1} \dots \pi_{\m_{s-3}} \cdot \pi_0^3+\dots)\ .
\eea
The extra factor $\frac{1}{s!}$ is taken for further convenience.
The auxiliary variables $y_0$ and $y^\m$ disappear form the expressions like $\hat\Phi^{(s)}(X,\Pi) \Psi^{(s)}(X,Y)$ which provide building blocks
for unconstrained Lagrangians to be discussed below.

In Sect. 4 we will consider fermionic massive higher spin fields. All the components will
carry an extra Dirac spinor index. In this case the definition of the conjugate composite field  (\ref{field}) should be modified so as to
include the conventional $\g_0$ standing on the right.

\vspace{0.5cm}

\noindent
{\bf 3. Massive spin-$s$ boson in flat space}\\

We begin by considering a quartet of fields $\Phi^{(s)},C^{(s-1)},D^{(s-2)},E^{(s-3)}$ and two Lagrange multipliers
$\Lambda^{(s-2)},\S^{(s-4)}$ in $\mathcal{D}+1$ dimensions which are subject to the following equations of motion \cite{bgk}
\bea\label{one}
&& P^2 \Phi^{(s)}-(YP) C^{(s-1)}+\frac{Y^2}{2} \Lambda^{(s-2)}=0, \qquad \quad C^{(s-1)}-(P\Pi)\Phi^{(s)} +(YP) D^{(s-2)}=0,
\nonumber\\[2pt]
&&
D^{(s-2)}-\frac{\Pi^2}{2} \Phi^{(s)} +(YP)E^{(s-3)}=0, \quad \qquad \quad \frac{\Pi^2}{2} D^{(s-2)} -(P\Pi)E^{(s-3)}=0,
\nonumber\\[2pt]\label{3}
&&
P^2 D^{(s-2)}-(P\Pi) C^{(s-1)}+\Lambda^{(s-2)}-Y^2 \S^{(s-4)}=0, \quad
\frac{1}{2} (P\Pi)\Lambda^{(s-2)}+(YP) \S^{(s-4)}=0\ .
\nonumber\\[2pt]
&&
\eea
This system holds invariant under the gauge transformation
\be\label{g}
\d \Phi^{(s)}=(YP) \Upsilon^{(s-1)}, \quad \d C^{(s-1)} =P^2 \Upsilon^{(s-1)},
\quad \d D^{(s-2)}=
(P\Pi)\Upsilon^{(s-1)}, \quad \d E^{(s-3)}=\frac{\Pi^2}{2} \Upsilon^{(s-1)}
\ee
with an unconstrained local parameter $\Upsilon^{(s-1)}$. As was demonstrated in \cite{bgk}, equations (\ref{one}) can be derived from a Lagrangian.
Moreover, $\Phi^{(s)}$ describes a massless spin--$s$ boson after eliminating the auxiliary fields
$C^{(s-1)},D^{(s-2)},E^{(s-3)}$. The Lagrange multipliers $\Lambda^{(s-2)},\S^{(s-4)}$ prove to vanish on--shell.

Let us apply the dimensional reduction mechanism outlined in the previous section to equations (\ref{one}). According to the
prescription (\ref{fiel}), each member of the quartet and each Lagrange multiplier yields a chain of fields in $\mathcal{D}$ dimensions
\bea
&&
\Phi^{(s)} \quad \quad ~ \rightarrow \quad (\phi^{(s)}, \phi^{(s-1)}, \dots, \phi)\ , \qquad ~ ~~
C^{(s-1)} \quad \rightarrow \quad (c^{(s-1)}, c^{(s-2)}, \dots, c)\ ,
\nonumber\\[2pt]
&&
D^{(s-2)} \quad~ \rightarrow \quad (d^{(s-2)}, d^{(s-3)}, \dots, d), \quad ~ \quad
E^{(s-3)} \quad \rightarrow \quad (e^{(s-3)}, e^{(s-4)}, \dots, e)\ ,
\nonumber\\[2pt]
&&
\Lambda^{(s-2)} \quad ~~\rightarrow \quad (\l^{(s-2)}, \l^{(s-3)}, \dots, \l), \quad \quad
\S^{(s-4)} \quad \rightarrow \quad (\s^{(s-4)}, \s^{(s-5)}, \dots, \s)\ .
\eea
The corresponding equations of motion for the components are derived from the set
\bea\label{eq1}
&&
(p^2-m^2) \Phi^{(s)}-(yp+imy_0) C^{(s-1)}+\frac{1}{2} (y^2+y_0^2) \Lambda^{(s-2)}=0,
\\[2pt]\label{eq2}
&&
C^{(s-1)}-(p\pi+im\pi_0)\Phi^{(s)} +(yp+imy_0) D^{(s-2)}=0,
\\[2pt]\label{eq3}
&&
D^{(s-2)}-\frac{1}{2} (\pi^2+\pi_0^2) \Phi^{(s)} +(yp+imy_0)E^{(s-3)}=0,
\\[2pt]\label{eq4}
&&\frac{1}{2} (\pi^2+\pi_0^2)D^{(s-2)} -(p\pi+im\pi_0)E^{(s-3)}=0,
\\[2pt]\label{eq5}
&&
(p^2-m^2) D^{(s-2)}-(p\pi+im\pi_0) C^{(s-1)}+\Lambda^{(s-2)}-(y^2+y_0^2) \S^{(s-4)}=0,
\\[2pt]\label{eq6}
&&
\frac{1}{2} (p\pi+im\pi_0)\Lambda^{(s-2)}+(yp+imy_0) \S^{(s-4)}=0\
\label{eq7}
\eea
by collecting the terms at each given power of $y_0$.

\begin{table}
\caption{The algebra of the Weyl--ordered operators quadratic in $(p,y,\pi)$}\label{table}
\begin{eqnarray}
\begin{array}{|l|r|r|r|r|r|c|}
\hline
[ {}{},{}{}] & p^2 & yp & p\pi & \frac{y^2}{2} & \frac{(y\pi+\pi y)}{2} & \frac{\pi^2}{2}
\\
\hline
p^2 & 0 & 0 & 0 & 0 &0 &0\\
\hline
(yp) & 0 & 0 & -p^2 & 0 & -yp & -p\pi\\
\hline
(p\pi) & 0 & p^2 & 0 & yp & p\pi & 0 \\
\hline
\frac{y^2}{2} & 0 & 0 & -yp & 0 & -y^2 & -\frac{(y\pi+\pi y)}{2}\\
\hline
\frac{(y\pi+\pi y)}{2} & 0 & yp & -p\pi & y^2 & 0 & -\pi^2\\
\hline
\frac{\pi^2}{2} & 0 & p\pi & 0 & \frac{(y\pi+\pi y)}{2} & \pi^2 & 0\\
\hline
\end{array}
\nonumber
\end{eqnarray}
\end{table}

\noindent

The gauge transformation (\ref{g})
takes the form
\bea\label{o}
&&
\d \Phi^{(s)}=(yp+imy_0)\Upsilon^{(s-1)}, \quad \quad ~ \d C^{(s-1)} =(p^2-m^2) \Upsilon^{(s-1)},
\nonumber\\[2pt]
&&
\d D^{(s-2)}=
(p\pi+im\pi_0)\Upsilon^{(s-1)}, \quad \d E^{(s-3)}=\frac{1}{2} (\pi^2+\pi_0^2) \Upsilon^{(s-1)}\ ,
\eea
where the gauge parameter $\Upsilon^{(s-1)}$ is to be understood as a composite object like in
(\ref{fiel})
\be
\Upsilon^{(s-1)} \quad \rightarrow \quad (\e^{(s-1)}, \e^{(s-2)}, \dots, \e)\ .
\ee
With the use of the commutators displayed in the table above one can readily verify that
equations (\ref{eq1})--(\ref{eq7}) are invariant under the transformation (\ref{o}).

From the first line in (\ref{o}) one concludes that the components $(\phi^{(s-1)}, \phi^{(s-2)}, \dots, \phi)\ $ entering
the composite field $\Phi^{(s)}$
can be gauged away and, as thus, are Stueckelberg fields. Let us demonstrate that the highest component $\phi^{(s)}$ describes a
free massive spin--$s$ bosonic field, while all the remaining fields vanish on--shell.

Multiplying (\ref{eq2}), (\ref{eq3}), (\ref{eq4}) by $-\frac{1}{2} (\pi^2+\pi_0^2)$, $(p\pi+im\pi_0)$, $(yp+imy_0)$, respectively, and
taking the sum, one gets
\be
(p^2-m^2)E^{(s-3)}-\frac{1}{2} (\pi^2+\pi_0^2) C^{(s-1)}=0\ .
\ee
This equation is then used to extract from (\ref{eq1})--(\ref{eq5}) the following restrictions on the Lagrange multipliers
\bea\label{L1}
&&
\frac{1}{4} (\pi^2+\pi_0^2) (y^2+y_0^2) \Lambda^{(s-2)}=\Lambda^{(s-2)}-(y^2+y_0^2) \S^{(s-4)}\ , \quad
\\[2pt]
&&
(\pi^2+\pi_0^2)(\Lambda^{(s-2)}-(y^2+y_0^2) \S^{(s-4)})=0\ .
\label{L2}
\eea

Before analyzing equations  (\ref{L1}) and (\ref{L2}), it is worth stopping for a moment to discuss
a technical issue. Consider the equation
\be\label{L3}
(\pi^2+\pi_0^2) (y^2+y_0^2) \Delta^{(s)}=0\ ,
\ee
where $\Delta^{(s)}$ is an arbitrary composite field as in (\ref{fiel}). Taking into account the identities
\bea\label{iden}
&&
[\frac{1}{2} (\pi^2+\pi_0^2),\frac{1}{2} (y^2+y_0^2)]=\frac{1}{2} (\mathcal{D}+1)+y\pi+y_0\pi_0\ ,
\nonumber\\[2pt]
&&
[\frac{1}{2} (\pi^2+\pi_0^2),\frac{1}{2} (\mathcal{D}+1)+y\pi+y_0\pi_0]=\pi^2+\pi_0^2\ ,
\eea
where $\mathcal{D}$ is the dimension of space--time, and the fact that $\Delta^{(s)}$ is a homogeneous function of degree $s$ in
$y^{\m_1} \dots y^{\m_{s-k}} {(y_0)}^k $
\be\label{Suppl}
(y\pi+y_0\pi_0) \Delta^{(s)} =s  \Delta^{(s)}\ ,
\ee
one concludes that $\Delta^{(s)}$ is
proportional to $(\pi^2+\pi_0^2)\Delta^{(s)}$. Acting  by the operator
$(\pi^2+\pi_0^2)$ on (\ref{L3}) and using (\ref{iden}), one can demonstrate that
$(\pi^2+\pi_0^2)\Delta^{(s)}$ is proportional to $(\pi^2+\pi_0^2)(\pi^2+\pi_0^2)\Delta^{(s)}$.
Clearly, this process can be continued. However, since $\Delta^{(s)}$ is a polynomial of the finite order in $y_0$ and $y^\m$, it
terminates at some step. Going backward one gets
\be\label{L4}
\Delta^{(s)}=0\ .
\ee

Let us turn back to  equations  (\ref{L1}), (\ref{L2}). Being combined, they imply
\be
(\pi^2+\pi_0^2)(\pi^2+\pi_0^2) (y^2+y_0^2) \Lambda^{(s-2)}=0\ \quad \rightarrow \quad (\pi^2+\pi_0^2)\Lambda^{(s-2)}=0\ .
\ee
Then the last line and the condition (\ref{L2}) yield
\be
\S^{(s-4)}=0\ ,
\ee
which on account of (\ref{L1}) gives
\be
\Lambda^{(s-2)}=0\ .
\ee
Thus, the Lagrange multipliers $\Lambda^{(s-2)}$ and $\S^{(s-4)}$ vanish on--shell.

As was mentioned above, the gauge symmetry (\ref{o}) allows one to gauge away all the components in the composite fields $\Phi^{(s)}$,
but for the highest component $\phi^{(s)}$. In our condensed notation the corresponding gauge choice reads
\be
\pi_0 \Phi^{(s)}=0\ .
\ee
Then successive multiplication of (\ref{eq1}) by $\pi_0$, $\pi_0^2$ and higher powers of $\pi_0$
allows one to relate $C^{(s-1)}$ to $\pi_0 C^{(s-1)}$,
$\pi_0 C^{(s-1)}$ to $\pi_0^2 C^{(s-1)}$ etc. However, since $C^{(s-1)}$ is a polynomial of the finite order in $y_0$, one concludes that
\be
C^{(s-1)}=0\ .
\ee
Clearly, equations (\ref{eq2}), (\ref{eq3}) can be treated in the same way and yield the result
\be
D^{(s-2)}=0, \qquad E^{(s-3)}=0\ .
\ee

Thus, in the gauge chosen, all the component fields entering the system (\ref{eq1})--(\ref{eq6}) vanish, but for $\phi^{(s)}$. The latter is
constrained to obey
the equations
\be
(p^2-m^2) \phi^{(s)}=0\ , \qquad (p\pi) \phi^{(s)}=0\ , \qquad \pi^2 \phi^{(s)}=0\ .
\ee
Eliminating the auxiliary variable $y^\m$, one gets the well known equations describing a free massive spin--$s$ boson in a
flat $\mathcal{D}$--dimensional space
\be\label{irrep}
(\Box-m^2)\phi_{\m_1 \dots \m_s}(x)=0\ , \qquad \partial^\n \phi_{\n \m_1 \dots \m_{s-1}}(x)=0\ , \qquad
{\phi^\n}_{\n \m_1 \dots \m_{s-2}}(x)=0\ .
\ee

Finally, we give an action functional which reproduces equations (\ref{eq1})--(\ref{eq6})
\bea\label{Act1}
&&
S=\int d^{\mathcal{D}} x \Bigl\{ \frac{1}{2} \hat\Phi^{(s)} (p^2-m^2) \Phi^{(s)}-
s\hat\Phi^{(s)}(yp+imy_0) C^{(s-1)} -\frac{1}{2} s\hat C^{(s-1)} C^{(s-1)}-
\nonumber\\[2pt]
&&
\quad \qquad
-s(s-1)\hat C^{(s-1)} (yp+imy_0) D^{(s-2)}-\frac{1}{2}s(s-1)\hat D^{(s-2)} (p^2-m^2) D^{(s-2)}+
\nonumber\\[2pt]
&&
\quad \qquad +\hat\Lambda^{(s-2)} \left( \frac{1}{2} (\pi^2+\pi_0^2) \Phi^{(s)}-s(s-1)D^{(s-2)}-\frac{1}{2}s(s-1)(s-2) (yp+imy_0) E^{(s-3)}\right)
\nonumber\\[2pt]
&&
\quad \qquad +\hat\Sigma^{(s-4)} \left(\frac{1}{2}s(s-1) (\pi^2+\pi_0^2) D^{(s-2)} -\frac{1}{2}s(s-1)(s-2)(p\pi+im\pi_0)E^{(s-3)}  \right)
\Bigr\}.
\eea
A formulation in terms of conventional tensor fields, i.e. components, can be easily read off from (\ref{Act1}) by substituting
the explicit form of the composite fields and taking the derivatives with respect to the auxiliary variables $y^\m$, $y_0$.
We would like to emphasize that, when passing to components,
all the coefficients in the action (\ref{Act1}) have a very simple form.
This is to be contrasted with the constrained formulation in
\cite{sh}.

That the action (\ref{Act1}) is invariant under the gauge transformation
\bea\label{O}
&&
\d \Phi^{(s)}=(yp+imy_0)\Upsilon^{(s-1)}, \quad \quad \qquad \qquad \d C^{(s-1)} =\frac{1}{s}(p^2-m^2) \Upsilon^{(s-1)},
\nonumber\\[2pt]
&&
\d D^{(s-2)}=\frac{1}{s(s-1)}
(p\pi+im\pi_0)\Upsilon^{(s-1)}, \quad \d E^{(s-3)}=\frac{1}{s(s-1)(s-2)} (\pi^2+\pi_0^2) \Upsilon^{(s-1)}\ ,
\nonumber\\[2pt]
&&
\eea
and yields (\ref{eq1})--(\ref{eq6}) as the equations of motion\footnote{To be more precise,
equations (\ref{eq1})--(\ref{eq6}) follow from the action (\ref{Act1}) after
the trivial field redefinition $s C^{(s-1)} ~\rightarrow ~C^{(s-1)}$, $s(s-1) \Lambda^{(s-2)} ~\rightarrow ~\Lambda^{(s-2)}$,
$s(s-1) D^{(s-2)} ~\rightarrow ~D^{(s-2)}$, $\frac{1}{2}s(s-1)(s-2) E^{(s-3)} ~\rightarrow ~E^{(s-3)}$,
$\frac{1}{2}s(s-1)(s-2)(s-3) \S^{(s-4)} ~\rightarrow ~\S^{(s-4)}$ .} is readily verified with the use of the Table 1 and the
identities
\bea\label{35}
&& \hat B^{(s-2)} \pi^2 A^{(s)}=s(s-1)\hat A^{(s)} y^2 B^{(s-2)}\ , \qquad \hat B^{(s-2)} \pi_0^2 A^{(s)}=s(s-1)\hat A^{(s)} y_0^2 B^{(s-2)}\ ,
\nonumber\\[4pt]
&& s \hat A^{(s)} (yp) B^{(s-1)}=-\hat B^{(s-1)} (p\pi) A^{(s)}\ , \qquad s \hat A^{(s)} y_0 B^{(s-1)}=-\hat B^{(s-1)} \pi_0 A^{(s)}\ ,
\nonumber\\[4pt]
&&
\hat A^{(s)} B^{(s)}=\hat B^{(s)} A^{(s)}\ .
\eea
The latter are valid for arbitrary composite fields $A$ and $B$ with {\it real} components. Notice that
the leftmost equation entering the second line in (\ref{35}) holds modulo a total derivative term which can be discarded under the integral
(\ref{Act1}).

Let us make a few comments on the structure
of the formulation (\ref{Act1}). First of all, the fields
in the Lagrangian and the gauge parameters do not obey any off-shell constraints, i.e. one has
a completely unconstrained formulation. Then, as is obvious
from equations (\ref{eq1})--(\ref{eq6}), the composite field
$C^{(s-1)}$ is purely auxiliary. It can be removed from the
consideration by solving the corresponding algebraic equation of
motion (\ref{eq2}). The collection of fields $(d^{(s-2)}, d^{(s-3)},
\dots, d)$ contained in the composite field $D^{(s-2)}$ is the
analogue of the auxiliary fields underlying the constrained
formulation by Singh and Hagen \cite{sh}. Solving (\ref{eq3}) for
$D^{(s-2)}$ is also feasible. This would lead to a higher
derivative formulation in the spirit of \cite{fms1}. The more general BRST
approach leads in this case to a Lagrangian which involves more
auxiliary fields \cite{bk1}.

Thus, the version containing two auxiliary composite
fields $D^{(s-2)}$, $E^{(s-3)}$ and two Lagrange multipliers $\Lambda^{(s-2)}$, $\S^{(s-4)}$ can be viewed as the minimal unconstrained
gauge invariant Lagrangian
formulation for a massive spin--$s$ boson in a flat $\mathcal{D}$--dimensional space\footnote{A possibility to describe massive higher spin bosons in
flat space in terms of a quartet of fields was discussed in \cite{tolya}. This formulation is given in terms of operators acting in a Fock space and
is applicable to the case $s\geq 4$. We thank M. Tsulaia for calling our attention to \cite{tolya}. }.

Before turning to fermionic fields, let us look at the system (\ref{eq1})--(\ref{eq6}) from a different angle.
Consider the first four equations in  (\ref{eq1})--(\ref{eq6}) with the Lagrange multipliers being discarded
\bea\label{A1}
&&
(p^2-m^2) \Phi^{(s)}-(yp+imy_0) C^{(s-1)}=0\ ,
\\[2pt]\label{A2}
&&
C^{(s-1)}-(p\pi+im\pi_0)\Phi^{(s)} +(yp+imy_0) D^{(s-2)}=0\ ,
\\[2pt]\label{A3}
&&
D^{(s-2)}-\frac{1}{2} (\pi^2+\pi_0^2) \Phi^{(s)} +(yp+imy_0)E^{(s-3)}=0\ ,
\eea
\bea\label{A4}
&&
\frac{1}{2} (\pi^2+\pi_0^2)D^{(s-2)} -(p\pi+im\pi_0)E^{(s-3)}=0\ .
\eea
It is easy to see that they are gauge invariant and describe a massive spin--$s$ boson.
The fifth equation
\bea\label{A5}
&&
(p^2-m^2) D^{(s-2)}-(p\pi+im\pi_0) C^{(s-1)}=0
\eea
proves to be a consequence of (\ref{A1})--(\ref{A4}). Equations (\ref{A1}) and (\ref{A2}) can be derived from a Lagrangian,
while, in order to get (\ref{A3}) and (\ref{A4}) from an action functional, one is forced to introduce two Lagrange multipliers
$\Lambda^{(s-2)}$, $\S^{(s-4)}$. Then the system (\ref{eq1})--(\ref{eq6}) can be viewed as an appropriate modification of (\ref{A1})--(\ref{A5})
such that what were previously identities among
(\ref{A1})--(\ref{A5}) turn into restrictions on the Lagrange multipliers which constrain them to vanish on--shell.
This method does not appeal to a massless theory living in $\mathcal{D}+1$ dimensions and proves to be particularly
convenient for describing fermions.

\vspace{0.5cm}

\noindent
{\bf 4. Massive spin-$s$ fermion in flat space}\\

Having constructed an unconstrained Lagrangian formulation for a massive higher spin boson in flat space, let us discuss massive higher spin
fermions. In this case, the dimensional reduction turns out to be less instructive because a naive reduction of Dirac matrices
from $\mathcal{D}+1$ to $\mathcal{D}$ dimensions does not yield a reasonable equation of motion. So, we choose to properly modify the analysis
in \cite{bgk}.

Consider a quartet of composite fields $\Psi_A^{(n)}$, $C_A^{(n-1)}$, $D_A^{(n-2)}$, $E_A^{(n-2)}$ which now carry an extra Dirac spinor index $A$.
We impose the following equations of motion\footnote{In what follows we keep spinor indices implicit. $\g^\m$ denote the standard Dirac
matrices which obey $\{\g^\m,\g^\n\}=-2\eta^{\m\n}$, $\eta^{\m\n}=\mbox{diag} (-,+,\dots,+)$. We use a representation in which ${(\g^0)}^{+}=\g^0$,
${(\g^0 \g^\m)}^{+}=\g^0\g^\m$.}
\bea\label{e1}
&&
(\g p-im)\Psi^{(n)}-(yp-im y_0) C^{(n-1)}=0\ ,
\\[2pt]\label{e2}
&&
C^{(n-1)}-(\g \pi +\pi_0) \Psi^{(n)}+(yp-im y_0) E^{(n-2)}=0\ ,
\\[2pt]\label{e3}
&&
D^{(n-2)}+\frac{1}{2} (\g p+im)E^{(n-2)}+\frac 12 (\g \pi -\pi_0) C^{(n-1)}=0\ ,
\\[2pt]\label{e4}
&&
(\g \pi +\pi_0) D^{(n-2)}-(p\pi-im \pi_0) E^{(n-2)}=0\ ,
\eea
which hold invariant under the gauge transformation
\bea\label{GT}
&&
\d \Psi^{(n)}=(yp -im y_0) \Upsilon^{(n-1)}\ , \qquad \quad \d C^{(n-1)}=(\g p -im) \Upsilon^{(n-1)}\ ,
\nonumber\\[2pt]
&&
\d D^{(n-2)}=(p\pi -im\pi_0) \Upsilon^{(n-1)}\ ,\qquad \d E^{(n-2)}=(\g\pi+\pi_0) \Upsilon^{(n-1)}\ .
\eea
As in the bosonic case, the gauge symmetry allows one to gauge away all component fields entering $\Psi^{(n)}$, but for the highest component which we
call $\psi^{(n)}(x)$. In our condensed notation the gauge choice reads
\be
\pi_0 \Psi^{(n)}=0\ .
\ee

Subsequent analysis goes along the same line as in the bosonic case. Acting by the operator $\pi_0$ on (\ref{e1}) one can relate
$C^{(n-1)}$ to $\pi_0 C^{(n-1)}$, $\pi_0 C^{(n-1)}$ to $\pi_0^2 C^{(n-1)}$ etc. which yields the result
\be
C^{(n-1)}=0\ .
\ee
Similarly, equation (\ref{e2}) gives
\be
E^{(n-2)}=0\ .
\ee
Then equation (\ref{e3}) constrains $D^{(n-2)}$ to vanish
\be
D^{(n-2)}=0\ .
\ee

Thus, in the gauge fixed form equations (\ref{e1})--(\ref{e4}) read
\be
(\g p-im)\psi^{(n)}=0\ , \quad (\g \pi) \psi^{(n)}=0\ \quad \rightarrow \quad (p\pi) \psi^{(n)}=0 \ ,
\ee
or, eliminating the auxiliary variable $y^\m$,
\be\label{final}
(\g^\n \partial_\n-im)\psi^{\m_1 \dots \m_n} (x)=0\ , \quad \g_\n \psi^{\n \m_1 \dots \m_{n-1}} (x)=0\ , \quad
\partial_\n \psi^{\n \m_1 \dots \m_{n-1}} (x)=0\ .
\ee
As is well known, equations (\ref{final}) describe a massive spin $s=n+\frac 12$ fermionic field in a flat
$\mathcal{D}$--dimensional space.

Notice that at this point equation (\ref{e4}) may seem redundant. However, it will come into a scene later on when we shall extend
(\ref{e1})--(\ref{e4}) so as to get a {\it Lagrangian} system of equations.

In order to construct an action functional reproducing equations (\ref{e1})--(\ref{e4}), let us introduce three Lagrange multipliers
(composite fields) $\Lambda^{(n-1)}$, $\S^{(n-2)}$, $\Omega^{(n-3)}$ which will accompany the constraints (\ref{e2})--(\ref{e4}) in a resulting Lagrangian.
It is assumed that the new fields are inert under the gauge transformation (\ref{GT}).

Then we consider two differential consequences of equations (\ref{e1})--(\ref{e4})
\bea\label{e5}
&&
(\g p+im) C^{(n-1)}+(p\pi -im \pi_0) \Psi^{(n)}-(yp-im y_0) D^{(n-2)}=0\ ,
\\[2pt]\label{e6}
&&
(\g p-im) D^{(n-2)}-(p\pi -im \pi_0) C^{(n-1)}=0\
\eea
and modify the resulting redundant system by including the Lagrange multipliers $\Lambda^{(n-1)}$, $\S^{(n-2)}$, $\Omega^{(n-3)}$ in the
following way
\bea\label{E1}
&&
(\g p-im)\Psi^{(n)}-(yp-im y_0) C^{(n-1)}+(\g y+y_0)\Lambda^{(n-1)}=0\ ,
\nonumber\\[2pt]
&&
(\g p+im) C^{(n-1)}+(p\pi -im \pi_0) \Psi^{(n)}-(yp-im y_0) D^{(n-2)}+\Lambda^{(n-1)}+\frac{1}{2} (\g y-y_0)\S^{(n-2)}=0\ ,
\nonumber\\[2pt]
&&
(\g p-im) D^{(n-2)}-(p\pi -im \pi_0) C^{(n-1)}+\S^{(n-2)}+(\g y+y_0)\Omega^{(n-3)}=0\ ,
\nonumber\\[2pt]
&&
C^{(n-1)}-(\g \pi +\pi_0) \Psi^{(n)}+(yp-im y_0) E^{(n-2)}=0\ ,
\nonumber\\[2pt]
&&
D^{(n-2)}+\frac{1}{2} (\g p+im)E^{(n-2)}+\frac 12 (\g \pi -\pi_0) C^{(n-1)}=0\ ,
\nonumber\\[2pt]
&&
(\g \pi +\pi_0) D^{(n-2)}-(p\pi-im \pi_0) E^{(n-2)}=0\ .
\eea
Notice that the constraints (\ref{e2})--(\ref{e4}) remain unchanged.

The idea behind the modification (\ref{E1}) is to convert what were previously identities among
(\ref{e1})--(\ref{e4}) and (\ref{e5}),(\ref{e6}) into restrictions on the Lagrange multipliers. Indeed, from
equations (\ref{E1}) one readily finds conditions which involve only the Lagrange multipliers
\bea\label{l1}
&&
\frac 12 (\g\pi-\pi_0)(\g y +y_0) \Lambda^{(n-1)}=-(\Lambda^{(n-1)}+\frac 12 (\g y-y_0) \S^{(n-2)}) \ ,
\\[2pt]\label{l2}
&&
\S^{(n-2)}+(\g y +y_0)\Omega^{(n-3)}=(\g\pi+\pi_0) (\Lambda^{(n-1)}+\frac 12 (\g y-y_0) \S^{(n-2)} )\ ,
\\[2pt]\label{l3}
&&
(\g\pi-\pi_0) (\S^{(n-2)}+(\g y +y_0)\Omega^{(n-3)})=0\ .
\eea
Taking into account the identities
\bea\label{l4}
&&
(\g\pi-\pi_0) (\g y+y_0)+(\g y-y_0) (\g\pi+\pi_0)=-(\mathcal{D}+1 +2(y\pi+y_0 \pi_0))\ ,
\nonumber\\[2pt]
&&
(\g\pi+\pi_0) (\g y-y_0)+(\g y+y_0) (\g\pi-\pi_0)=-(\mathcal{D}+1 +2(y\pi+y_0 \pi_0))\ ,
\nonumber\\[2pt]
&&
(\g\pi) (\g y)+(\g y) (\g\pi)=-2 (y\pi) -\mathcal{D}
\eea
and the homogeneity condition (\ref{Suppl}) which is valid for an arbitrary composite field, one can demonstrate that all the
Lagrange multipliers vanish on--shell.

The proof is similar to the bosonic case and goes as follows. Acting by the operator $(\g\pi-\pi_0)(\g\pi+\pi_0)$ on (\ref{l1})
and taking into account (\ref{l2}), (\ref{l3}), (\ref{l4}) one gets
\be
(\g\pi-\pi_0)(\g\pi+\pi_0)(\g\pi-\pi_0)(\g y +y_0) \Lambda^{(n-1)}=0 \quad \rightarrow \quad (\g\pi-\pi_0)(\g\pi+\pi_0)\Lambda^{(n-1)}=0\ .
\ee
The last line along with (\ref{l2}), (\ref{l3}) yields
\be
(\g\pi-\pi_0)(\g\pi+\pi_0)(\g y-y_0) \S^{(n-2)}=0 \quad \rightarrow \quad (\g\pi-\pi_0)\S^{(n-2)}=0\
\ee
which, in view of (\ref{l3}), constrains $\Omega^{(n-3)}$ to vanish
\be
(\g\pi-\pi_0)(\g y +y_0)\Omega^{(n-3)}=0 \quad \rightarrow \quad \Omega^{(n-3)}=0 \ .
\ee
At this point (\ref{l2}) allows one to express $\S^{(n-2)}$ in terms of $(\g\pi+\pi_0)\Lambda^{(n-1)}$ which after substitution in
(\ref{l1}) yields
\be
\Lambda^{(n-1)}=0 \quad \rightarrow \quad \S^{(n-2)}=0\ .
\ee
Thus, the extended system (\ref{E1}) is equivalent to equations (\ref{e1})--(\ref{e4}) and, hence, describes a massive spin
$s=n+\frac 12$ fermionic field in flat space.

The advantage of the extended version is that it can be derived from the
action functional\footnote{To be more precise, equations (\ref{E1}) arise from the action after the field redefinition
 $n C^{(n-1)} ~\rightarrow ~C^{(n-1)}$, $n(n-1) D^{(n-2)} ~\rightarrow ~D^{(n-2)}$, $n(n-1) E^{(n-2)} ~\rightarrow ~E^{(n-2)}$,
 $n \Lambda^{(n-1)} ~\rightarrow ~\Lambda^{(n-1)}$, $n(n-1) \S^{(n-2)} ~\rightarrow ~\S^{(n-2)}$, $n(n-1)(n-2) \Omega^{(n-3)}
 ~\rightarrow ~\Omega^{(n-3)}$.}
\bea\label{FFIN}
&&
S=\int dx^{\mathcal{D}} \Bigl\{i\hat{\bar\Psi}^{(n)}((\g p-im)\Psi^{(n)}-n(yp-im y_0) C^{(n-1)}+n(\g y+y_0)\Lambda^{(n-1)})-
\nonumber\\[2pt]
&& \qquad
-i\hat{\bar C}^{(n-1)}(n(\g p+im) C^{(n-1)}+(p\pi -im \pi_0) \Psi^{(n)}-n(n-1)(yp-im y_0) D^{(n-2)}+
\nonumber\\[2pt]
&& \qquad +
n\Lambda^{(n-1)}+\frac 12 n(n-1)(\g y-y_0)\S^{(n-2)} )-i\hat{\bar D}^{(n-2)}(n(n-1)(\g p-im) D^{(n-2)}-
\nonumber\\[2pt]
&&
\qquad -n(p\pi -im \pi_0) C^{(n-1)}+n(n-1)\S^{(n-2)}+n(n-1)(n-2)(\g y+y_0) \Omega^{(n-3)})+
\nonumber\\[2pt]
&& \qquad +
i\hat{\bar \Lambda}^{(n-1)} (nC^{(n-1)}-(\g \pi +\pi_0) \Psi^{(n)}+n(n-1)(yp-im y_0) E^{(n-2)})+
\nonumber\\[2pt]
&& \qquad +
i\hat{\bar \S}^{(n-2)}(n(n-1)D^{(n-2)}+\frac{1}{2} n(n-1)(\g p+im)E^{(n-2)}+\frac 12 n (\g \pi -\pi_0) C^{(n-1)})+
\nonumber\\[2pt]
&& \qquad +
i\hat{\bar \Omega}^{(n-3)} (n(n-1)(\g \pi +\pi_0) D^{(n-2)}-n(n-1)(p\pi-im \pi_0) E^{(n-2)})+
\nonumber\\[2pt]
&& \qquad +
i\hat{\bar E}^{(n-2)} (\frac{1}{2} n(n-1) (\g p +im) \S^{(n-2)}+n(p\pi-im\pi_0)\Lambda^{(n-1)}-
\nonumber\\[2pt]
&& \qquad
-n(n-1)(n-2)(yp-imy_0)\Omega^{(n-3)})
 \Bigl\}\ .
\eea
A formulation in terms of
conventional spin--tensors, i.e. components, can be easily read off
from (\ref{FFIN}) by substituting the explicit form of the composite
fields and taking the derivatives with respect to the auxiliary
variables $y^\m$, $y_0$.

That the action is real is readily verified with the use of the identities
\bea\label{B}
&&
{(n \hat{\bar A} ^{(n)}(yp) B^{(n-1)})}^{\dagger}=-\hat{\bar B} ^{(n-1)}(p\pi) A^{(n)}\ ,\quad
{(n \hat{\bar A} ^{(n)} y_0 B^{(n-1)})}^{\dagger}=\hat{\bar B} ^{(n-1)} \pi_0 A^{(n)}\ ,
\nonumber\\[2pt]
&&
{(n \hat{\bar A} ^{(n)}(\g y) B^{(n-1)})}^{\dagger}=\hat{\bar B} ^{(n-1)}(\g\pi) A^{(n)}\ , \quad
{(\hat{\bar A} ^{(n)} B^{(n)})}^{\dagger}=\hat{\bar B} ^{(n)} A^{(n)}\ .
\eea
The leftmost equation entering the first line in (\ref{B}) holds modulo a total derivative term which can be discarded under the integral (\ref{FFIN}).
The gauge transformation leaving (\ref{FFIN}) invariant reads
\bea
&&
\d \Psi^{(n)}=(yp -im y_0) \Upsilon^{(n-1)}\ , \qquad \quad \qquad \quad \d C^{(n-1)}=\frac 1n (\g p -im) \Upsilon^{(n-1)}\ ,
\nonumber\\[2pt]
&&
\d D^{(n-2)}=\frac{1}{n(n-1)}(p\pi -im\pi_0) \Upsilon^{(n-1)}\ ,\quad \d E^{(n-2)}=\frac{1}{n(n-1)}(\g\pi+\pi_0) \Upsilon^{(n-1)}\ .
\eea

Analogously to the bosonic case, we obtained a formulation in terms
of unconstrained fields and gauge parameters with very simple number coefficients in the
Lagrangian. Notice, however, that, in contrast to the bosonic case,
elimination of the auxiliary composite field $C^{(n-1)}$ would lead
to higher derivative terms. Thus, the formulation above can be
viewed as the minimal unconstrained gauge invariant Lagrangian
formulation for a massive spin--$(n+\frac 12)$ fermion in a flat
$\mathcal{D}$--dimensional space.

\vspace{0.5cm}

\noindent
{\bf 5. Conclusion}\\

To summarize, in this work we generalized the quartet unconstrained
description of massless higher spin fields \cite{bgk} to the case of
massive higher spin fields in a flat space of arbitrary dimension.
Our Lagrangian formulation is given in terms of unconstrained fields
and gauge parameters and has an
easy-to-handle form for an arbitrary value of spin. It is
local, free from higher derivative terms and involves a minimal
number of auxiliary fields needed for an unconstrained gauge
invariant description of a free massive spin--$s$ field. Explicit evaluation
of the number coefficients in the Lagrangian is very simple and
does not require a complicated procedure as in
\cite{sh}.

The quartet formulation occupies an intermediate position
between the general BRST formulation of \cite{bk1,bk2} and the
geometric approach of \cite{fr,fms1} unifying in a nice way their
advantages and avoiding their disadvantages. It is natural to
expect that the quartet formulation can be obtained from the
BRST method by partial gauge fixing and eliminating some of
the auxiliary fields.

Let us mention a few possible developments of the present work.
First of all, it would be interesting to extend the present
consideration to the case of a massive spin--$s$ particle
propagating on anti de Sitter background. Then it is interesting to study whether
the quartet unconstrained massive gauge theory in anti de Sitter space
can be obtained  by means of
the dimensional degression discussed recently in \cite{av}. It is also interesting to
generalize the analysis to the case of mixed--symmetry tensor fields
and to construct supersymmetric generalizations.

\vspace{0.5cm}

\noindent{\bf Acknowledgements}\\

\noindent
We thank D. Francia, V. Krykhtin, A. Sagnotti and M. Tsulaia for useful comments. The research was supported by RF Presidential grants
MD-2590.2008.2, NS-2553.2008.2 and RFBR grant 08-02-90490-Ukr.

\end{document}